\documentclass{article}
\usepackage{amsfonts}
\usepackage{amsmath}
\usepackage{makeidx}
\usepackage{eurosym}
\usepackage{amssymb}
\usepackage{graphicx}
\usepackage{caption}
\usepackage{wrapfig}
\usepackage{subfig}
\usepackage{rotating}
\usepackage{natbib}
\usepackage{appendix}
\newcommand{\bm}{\bibitem}
\newcommand{\be}{\begin{equation}}
\newcommand{\ee}{\end{equation}}
\newcommand{\bea}{\begin{eqnarray}}
\newcommand{\bes}{\begin{subequations}}
\newcommand{\ees}{\end{subequations}}
\newcommand{\bgt}{\begin{gather}}
\newcommand{\egt}{\begin{gather}}
\newcommand{\eea}{\end{eqnarray}}
\newcommand{\beaa}{\begin{eqnarray*}}
\newcommand{\eeaa}{\end{eqnarray*}}

\newcommand{\EE}{{\mathbb E}}
\newcommand{\PP}{{\mathbb P}}
\newcommand{\RR}{{\mathbb R}}

\title{SPX, VIX and scale-invariant LSV\footnote{Local Stochastic Volatility} }
\author{Alexander Lipton \& Adil Reghai }
\date{First version November 2022 / This version February 2023}

\begin{document}

\maketitle

\begin{abstract}

Local Stochastic Volatility (LSV) models have been used for pricing and hedging derivatives positions for over twenty years. An enormous body of literature covers analytical and numerical techniques for calibrating the model to market data. However, the literature misses a potent approach commonly used in physics and works with absolute (dimensional) variables rather than with relative (non-dimensional) ones. While model parameters defined in absolute terms are counter-intuitive for trading desks and tend to be heavily time-dependent, relative parameters are intuitive and stable, making it easy to steer the model adequately and consistently with its Profit and Loss (PnL) explanation power. We propose a specification that first explores historical data and uses physically well-defined relative quantities to design the model. We then develop an efficient hybrid method to price derivatives under this specification. We also show how our method can be used for robust scenario generation purposes - an important risk management task vital for buy-side firms.\footnote{The authors would like to thank Prof. Marcos Lopez de Prado and Dr. Vincent Davy Zoonekynd for valuable comments.}

\end{abstract}

\section{Introduction}

Modern financial engineering started with the Black \& Scholes model developed; see \cite{Black73}. This model has created a consensus for pricing the simplest derivative options, vanilla ones. Black \& Scholes implied volatility is a shared concept that any market practitioner can infer from market prices. It is the solid pillar behind the volatility smile description.

However, the Black \& Scholes model has severe limitations when pricing advanced derivatives, especially those with changing second-order exposures; see, e.g., \cite{Reghai15} when it comes to the gamma exposures. Following the first principles of derivative valuation, the explanation of PnL or replication is the primary judge of the quality of a model; see. e.g., \cite{Lipton02}. Thus, practitioners needed a model that could consistently price those derivatives with the volatility smile. \cite{Bick93}, \cite{Dupire94}, and \cite{Derman96} developed such a model through the introduction of local volatility. This model solved the pricing issue in the presence of the smile, particularly the changing gamma exposure over time and spot value. However, despite its wide usage by the industry, the local volatility model only included the static cost of hedging the smile. In theory, the model accounts only for the introduction of the different vanillas at the inception of the trade. In practice, traders, risk managers, and quants all realized quickly that during the life of the trade, an additional systematic cost arises when the trader readjusts her position. Nevertheless, the readjustment costs are not adequately considered under the local volatility model. Therefore, the industry needed models consistent with the static and dynamic cost of hedging. 

The solution for this problem was introduced long ago and labeled the Local Stochastic Volatility (LSV) model of \cite{Jex99} and \cite{Lipton02}. In theory, the model was perfectly adequate. However, in practice, both the local volatility and the LSV had specifications pitfalls that generated recurrent problems when applied on the industrial scale. In particular, they were based on absolute levels of the spot. Some practitioners partially tackled the problem by normalizing with the forward values to circumvent this issue. However, little information was documented, resulting in arbitrary choices and significant differences from one bank to the other. Accordingly, although prices were marginally impacted, the resulting hedges were not necessarily correct.

In \cite{Hobson98}, an exciting class of path-dependent local volatility models was proposed. A specific version was examined with the volatility depending on the difference between the current price and an exponential average of past prices. This property is appealing to traders as it reflects the perception that large movements of the asset price in the past tend to forecast higher future volatility. This model exhibits a wide variety of smiles and skews. Lipton emphasized that all meaningful models have to be scale-invariant (dimensionless); see \cite{Lipton99, Lipton01, Lipton17}. Lipton recommended building physical models depending on functions of dimensionless arguments of the time and spot parameters $t$ and $S_t$ and introduced the general form of such a model based on a kernel. In \cite{Guyon14, Guyon22}, the same observation was made regarding local volatility setup, showing how to capture prominent historical volatility patterns using path-dependent local volatility. 

\cite{Hagan02} present an analytically tractable formula that is scale-invariant and has both local and stochastic volatility. However, the local volatility part, being parametric and rigid, does not permit fitting the whole volatility surface. 

We need to keep several typical trading conundrums in mind to design a proper approach to the problem at hand.   
\begin{itemize}
    \item Traders know very well that path dependency strongly impacts the volatility dynamic. It is an essential factor in their PnL explanation for the cost of hedging. Consider the Spx index. The implied volatility surface when the spot level is at 4000 depends on its previous levels. On the one hand, if the spot went down before, then the implied volatility surface would be higher, reflecting an elevated level of risk. On the other hand, if the spot comes from lower levels, then the implied volatility is certainly lower as the market is pricing a momentum movement. 
    \item Traders think in relative terms, and to be consistent with the industry, the volatility modeling must rely on a dimensionless approach. 
    \item Traders operate on multiple time scales. They infer the cost of future volatility hedging through the recent evaluation of the market movement. They know that different players (institutional investors, asset managers, day traders, and statistical arbitrageurs) all impact market volatility.
    \item This set of models is vital for correctly determining the volatility dynamic, which is essential for hedging vanilla options and pricing path dependant options such as barrier or autocall options.
\end{itemize}

Our paper aims to provide an in-depth analysis of the most traded assets, SPX and VIX. Section 2 presents the SDE governing the LSV model. Section 3 explores data and presents the critical link between SPX and VIX. It shows that the scale-invariant specification gives excellent results both in-sample and out-of-sample. Section 4 presents different techniques for pricing derivatives, some of which are highly efficient and useful for production purposes. In particular, the impact on pricing classical path-dependent options is reasonably high. Section 5 briefly deals with scenario generation and involves creating hypothetical scenarios to simulate different realistic market conditions. Finally, Section 6 concludes.

The paper provides a comprehensive understanding of the financial market and is valuable for financial analysts, traders, and risk managers.

\section{The process}

We are interested by the process of the form:
\bea
    \frac{d S_{t}}{S_{t}} &=& (r-q)dt + \sigma_{loc}  e^{Y_t} (\rho dW_t + \sqrt{1-\rho^2} dB_t ),\\
    dY_t & = & -{\kappa} Y_t dt + \nu dW_t, 
\eea
where $r,q$ are the risk neutral rate and dividend yield,
\bea
\sigma_{loc} = f\left({t \over \bar{t}}, \frac{S_t}{ \int_{-\infty}^t \phi(t,u) S_udu}\right),
\eea
where $\bar{t}$ is a representative timescale, and $\phi$ is a suitable averaging kernel, for example 
\bea
\phi(t,u)=\kappa e^{-\kappa (t-u)}.
\eea

The scale invariant index $I_t=\frac{S_t}{ \int_{-\infty}^t \phi(t,u) S_udu}$ satisfies the following stochastic differential equation (SDE):

\bea
    \frac{dI_t}{I_t} &=& ((r-q)-\phi(t,t)I_t) dt + \sigma_{loc}  e^{Y_t} (\rho dW_t + \sqrt{1-\rho^2} dB_t ).
\eea

The natural scale for this index is the unity, and a deviation from it can be seen as a shock or a surprise and therefore generates a different value for the risk. Risk, in our case, is a forward-looking measure that the VIX represents.

\begin{figure}[!h]
    \centering
    \includegraphics[scale=0.5]{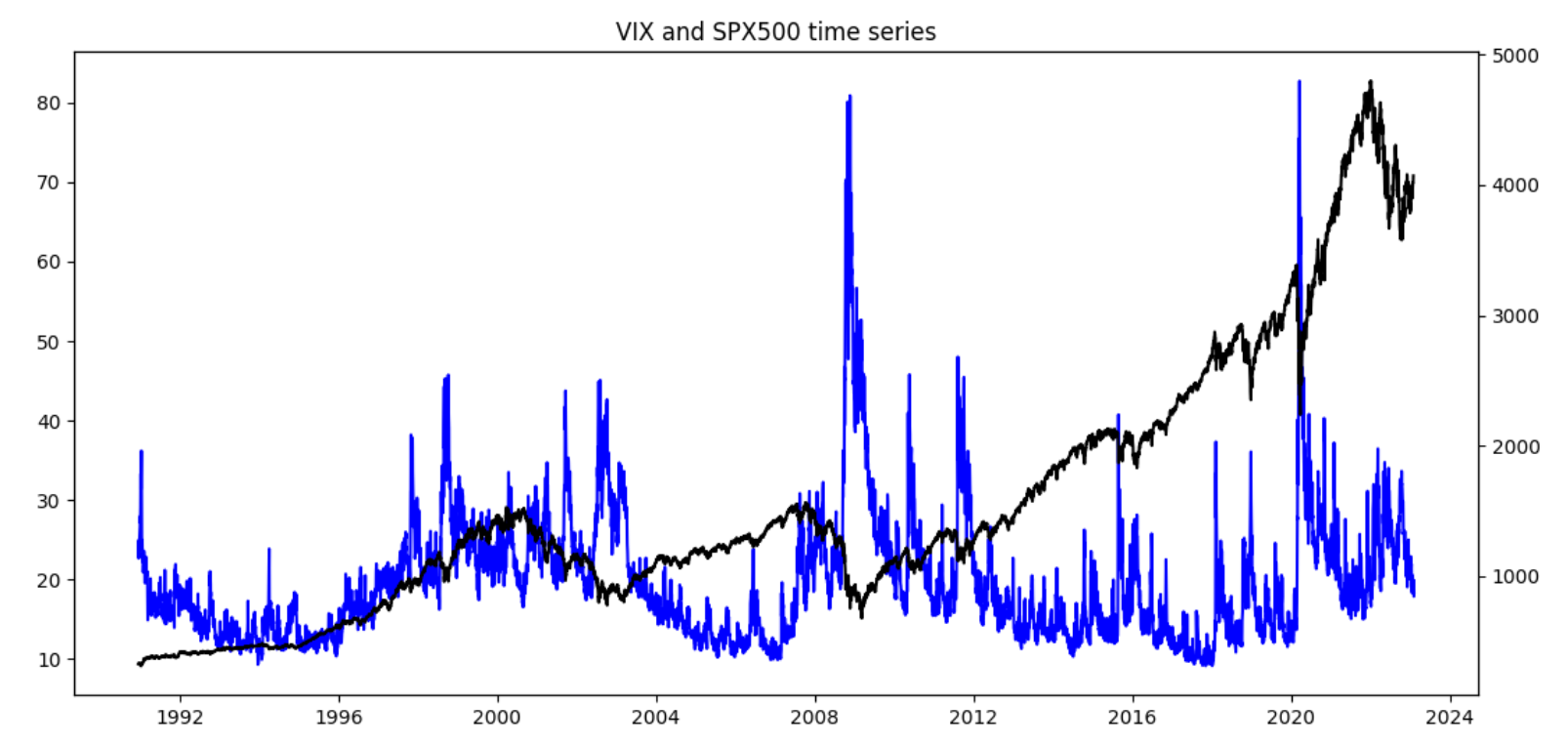}
    \caption{VIX and SPX as functions of time. Own graphics. Source : Factset.}
    \label{fig:SPX VIX}
\end{figure}

\section{ Data exploration}

In this section, we will explore historical data. We shall observe the historical paths for the SPX and VIX and qualitatively determine the type of links between these two quantities. In particular, we shall see that describing the VIX as a function of the absolute level of the SPX is not straightforward. Also, using a moving average kernel, we observe that this creates more data structure. Finally, we shall seek an optimal kernel designed to maximize the fit of the VIX as a function of the dimensionless quantity of the path.

\subsection{VIX as a function of the spot}
We consider a long period of time starting from the beginning of the 1990 till the end of 2022.

\begin{figure}
    \centering
    \includegraphics[scale=0.5]{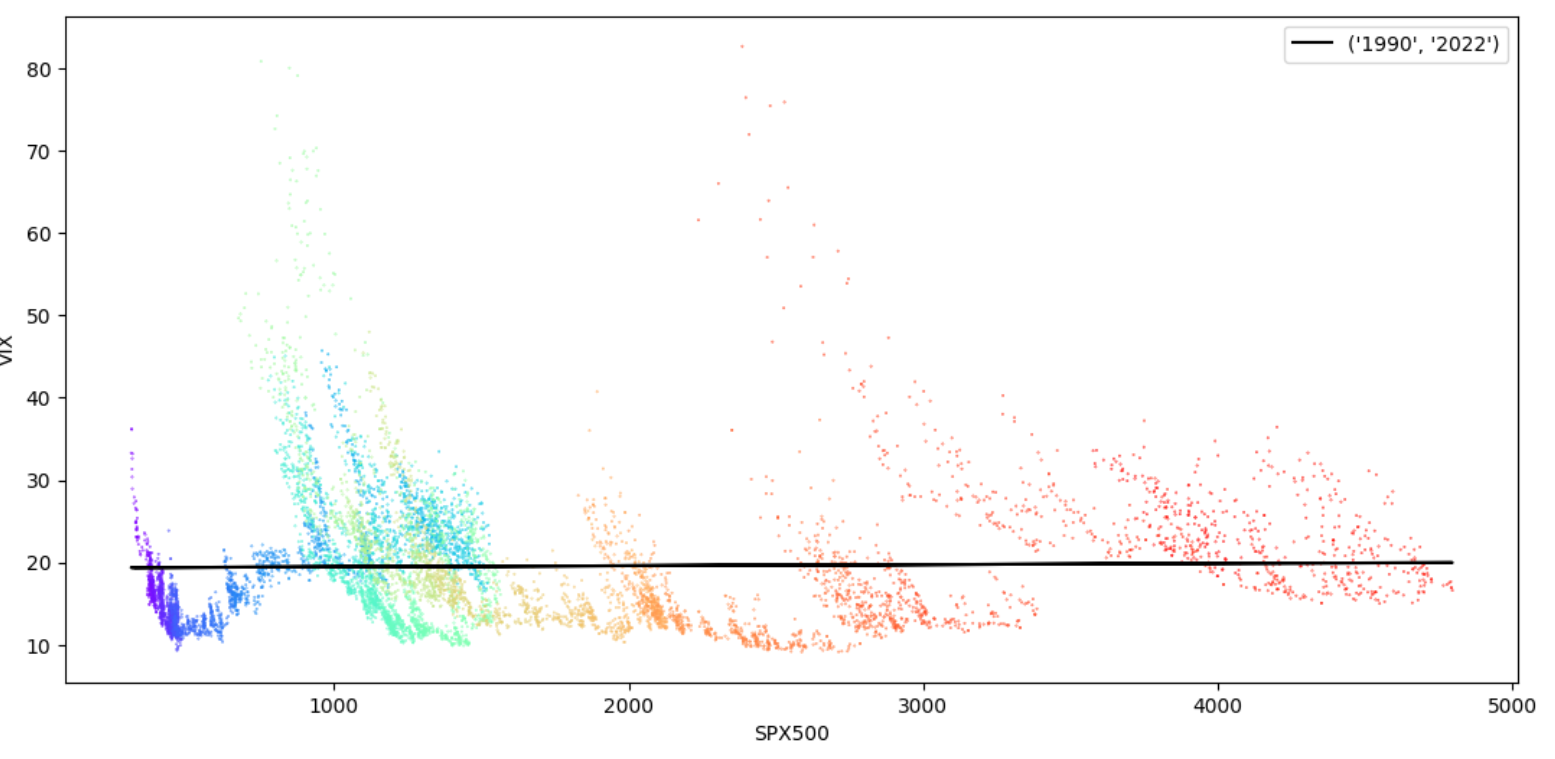}
    \caption{Global Fit. Own graphics.}
    \label{fig:globalfit}
\end{figure}

\begin{figure}
    \centering
    \includegraphics[scale=0.5]{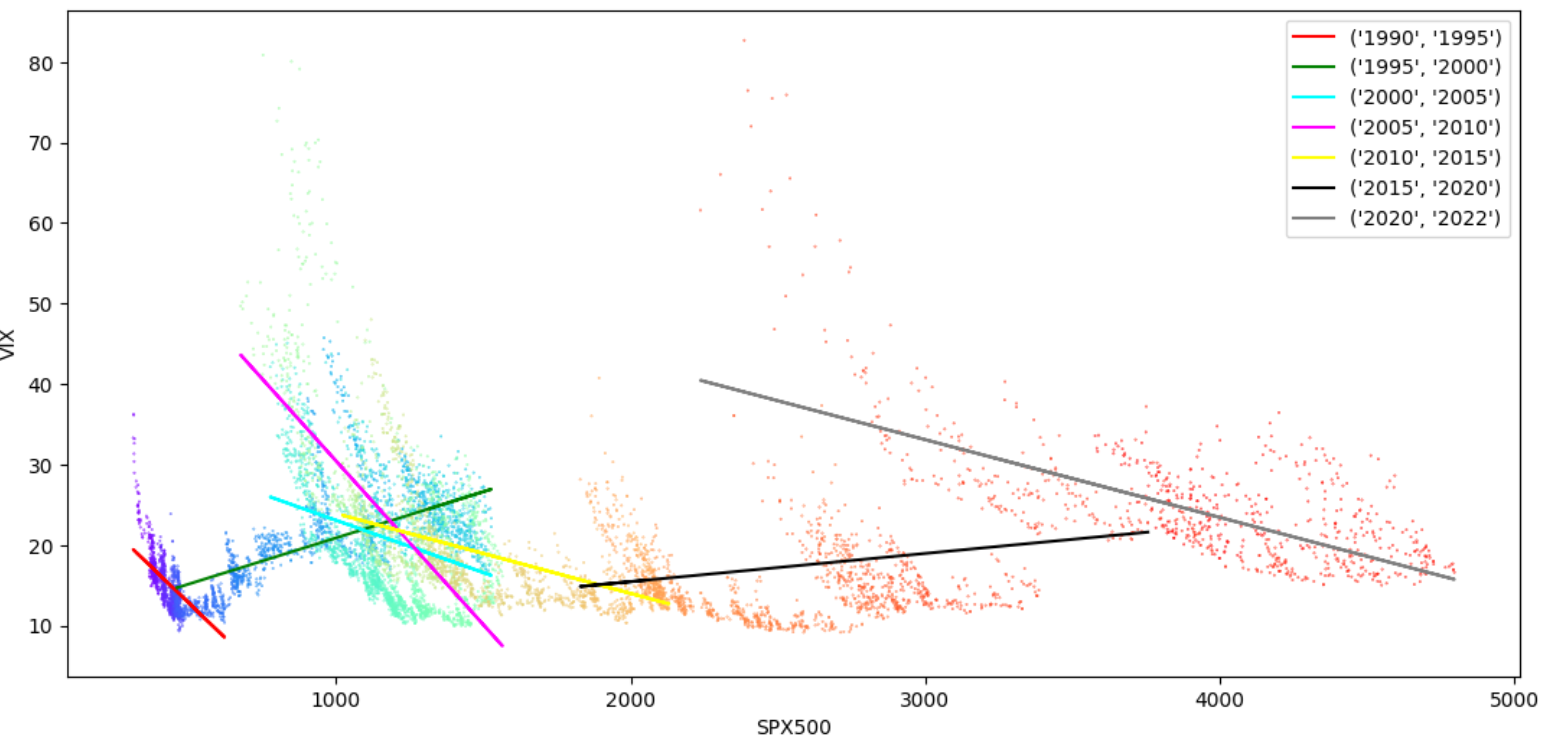}
    \caption{Local Fit per period of 5 years. Own graphics.}
    \label{fig:localfit}
\end{figure}

\begin{table}[!h]
    \centering
    \begin{tabular}{|c|c|c|c|}
    \hline
    Period & Intercept & Slope & R2 Score  \\ \hline
    [1990-1995] & 30.4 & -0.0352 &  42.78  \\ \hline
    [1995-2000] & 9.4 & 0.0115 & 38.34  \\ \hline
    [2000-2005] & 36.06 & -0.013 & 10.34  \\ \hline
    [2005-2010]  & 71.14 & -0.0407 & 44.02  \\ \hline
    [2010-2015] & 33.85 & -0.0099 &  32.67 \\ \hline
    [2015-2020] & 8.44 & 0.0035 & 3.73  \\ \hline
    [2020-    ] & 62.06 & -0.0097 &  37.27 \\ \hline
    \end{tabular}
    \caption{Score per period of 5 years}
    \label{tab:scores per period}
\end{table}

In Figures \ref{fig:SPX VIX}, \ref{fig:globalfit}, we can observe that the VIX as a function of the absolute spot level shows little structure as expected because traders recognize many skewed regimes associated with different periods. Each period corresponds to some characteristic level of spot. In Figure \ref{fig:localfit}, we observe different local behaviour of the volatility relatively to the absolute level of the SPX. In particular, one can see that for a level of around 3500 there is hysteresis, i.e. the volatility is either increasing or decreasing depending on the period under scrutiny. The levels of R2 scores are shown in Table \ref{tab:scores per period}. 

\begin{figure}[!h]
    \centering
    \includegraphics[scale=0.9]{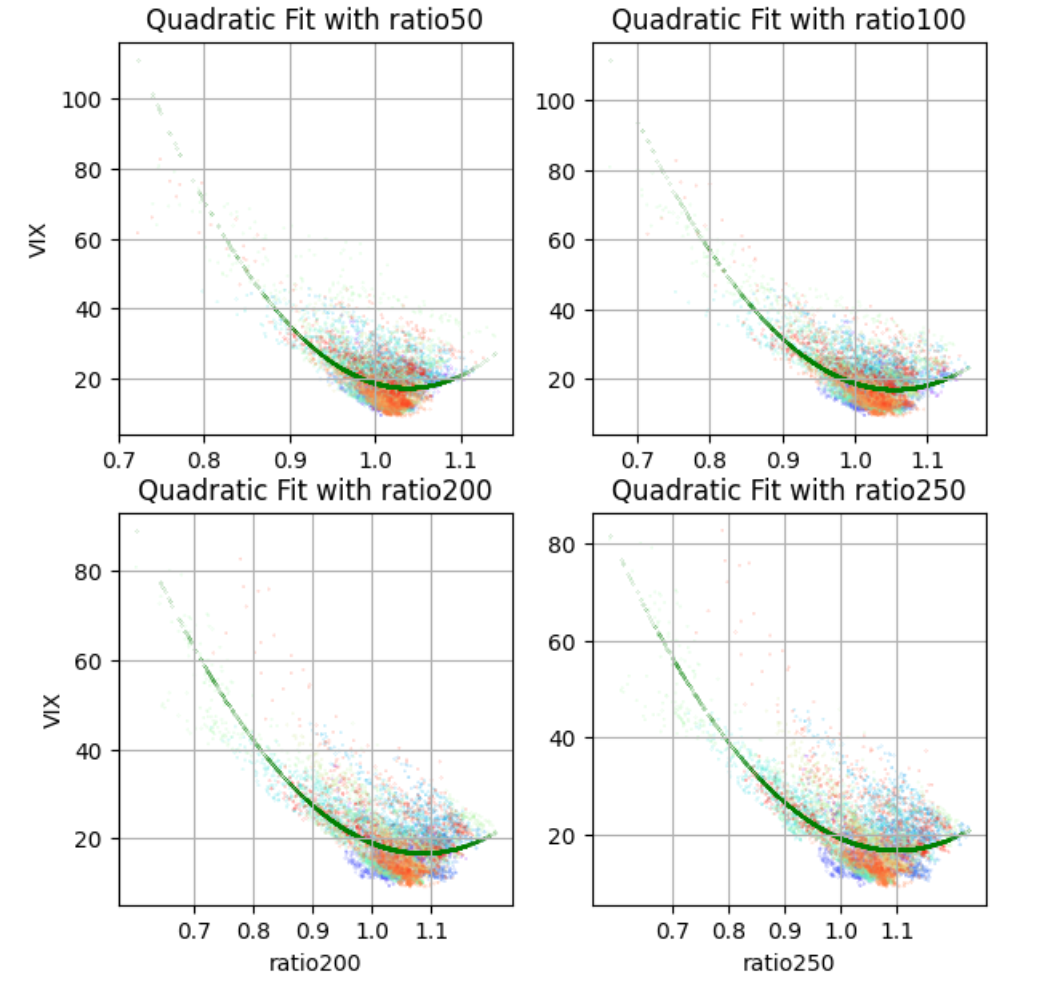}
    \caption{Fit using natural trading scales. Own graphics.}
    \label{fig:ewma50}
\end{figure}

In Figure \ref{fig:ewma50}, we try different classical moving averages that are classical in the trading industry. Each moving average is a particular, in equation (\ref{eq:kernel}), a kernel with equal weights for each spot within the averaging window is shown. More precisely, for a lag of $n$ days, the kernel is given by:

\begin{equation}
    \phi(t,u) = \frac{1}{n} {1}_{t-u \leq n}.    
    \label{eq:kernel}
\end{equation}

We examine different lags based on trading rationals that are exposed below.

\begin{itemize}
    \item 50 days: The 50-day moving average is a reliable technical indicator used by several investors to analyze price trends. It is a security's average closing price over the previous 50 days. The 50-day moving average is popular because it is a realistic and effective trend indicator in the stock market.
    \item 100 days: A moving average of 100 days helps investors see how the stock has performed over 20 weeks and find the price trend if it is upward or downward, which gives them a sense of the market sentiment as well. 
    \item 200 days: The 200-day moving average is perceived as the dividing line between a technically healthy stock and one that is not. Furthermore, the percentage of stocks above their 200-day moving average helps determine the market's overall health.
    \item 250 days: The 250 period moving average is popular on the daily chart since it describes one year of the price action (one year has roughly 250 trading days).
\end{itemize}

\begin{table}
    \centering
    \begin{tabular}{|c|c|c|c|c|}
    \hline
    Method & a & b & c & score \\ \hline
    50 days & 0 & -1966.13 &  947.69 & 44.1   \\ \hline
    100 days & 0 & -1302. & 618.  &  55.29    \\ \hline
    200 days & 0 & -676 & 311.  &   56.24    \\ \hline
    250 days  & 0 & -548. & 250.  & 54.83     \\ \hline
    \end{tabular}
    \caption{$Vix(t) = a+bx+cx^2$}
    \label{tab:natural fit}
\end{table}

Note that in Table \ref{tab:natural fit} that the constant coefficient is not loaded, which is natural since we are regressing two non-dimensional quantities. Also, note that the ratio between the slope coefficient is negative for all time scales, as expected. Finally, the ratio of the slope coefficient and the curvature one are all of the same levels, i.e., the level of the average VIX value. For the last experiment, see equations (\ref{equation:optim1}, \ref{equation:optim2}, \ref{equation:optim3}) we perform an optimization described as follows. We choose a length $n$ for the kernel. We assume that it is a stationary kernel with positive weights.

\begin{equation}
\min_{\phi_1,...,\phi_n,a,b,c} \frac{1}{T-n+1} \sum_{t=n}^T (VIX_t-(a+bI+cI^2))^2    
\label{equation:optim1}
\end{equation}

where
\begin{equation}
I = \frac{S_t}{\phi_1 S_{t-n}+...+\phi_nS_{t-1}},
\label{equation:optim2}
\end{equation}

subject to:
\begin{eqnarray}
   \phi_1+...+\phi_n & = & 1 \\
   \phi_i & \geq & 0 \quad \forall i
   \label{equation:optim3}
\end{eqnarray}

We use a classic reparametrization of the weights, as in equation (\ref{equation:optim4}),  to perform a non-constrained optimization. More precisely, we search for $\psi_i$ such that:

\begin{equation}
\phi_i = \frac{e^{\psi_i}}{\sum_j e^{\psi_j}},    
\label{equation:optim4}
\end{equation}
Now $\psi_i, \forall i$ are free from constraints.

\subsection{In-sample results}

Below we perform an in-sample study where we identify the functional form for the period [1990-2010].

\begin{table}[!h]
    \centering
    \begin{tabular}{|c|c|c|c|c|}
    \hline
    Method & a & b & c & R2 score \\ \hline
    Optimised 50 days & 72.12 & 3.96 &  -56.32 & 82.77   \\ \hline
    Optimised 100 days & 92.25 & 2.49 & -74.2 &  86.73    \\ \hline
    Optimised 200 days & 66.18 & 4.25 & -49.17 & 88.07    \\ \hline
    Optimised 250 days  & 75.02 & 3.47 & -57.13 & 88.55     \\ \hline    
    \end{tabular}
    \caption{$Vix(t) = a+bx+cx^2$}
    \label{tab:quadraticoptimisedfit}
\end{table}
Table \ref{tab:quadraticoptimisedfit} demonstrates that the quality of the $R^2$ score has improved significantly to up to 88\%. We shall measure the fit quality for the out-of-sample period in the next section.

\subsection{Out-of-sample results}

This section measures the score of the previously calibrated models for the out-of-sample period [2010-2022]. The corresponding results are shown in Table \ref{tab:outofsamplefit}.

\begin{table}[!h]
    \centering
    \begin{tabular}{|c|c|}
    \hline
    Method & R2 score \\ \hline
    Optimised 50 days & 84.09   \\ \hline
    Optimised 100 days & 87.7    \\ \hline
    Optimised 200 days &  88.4    \\ \hline
    Optimised 250 days  &  88.53     \\ \hline
    \end{tabular}
    \caption{$R^2$ score using optimised weights.}
    \label{tab:outofsamplefit}
\end{table}

\subsection{Optimised weights}

\begin{figure}[!h]
    \centering
    \includegraphics[scale=0.79]{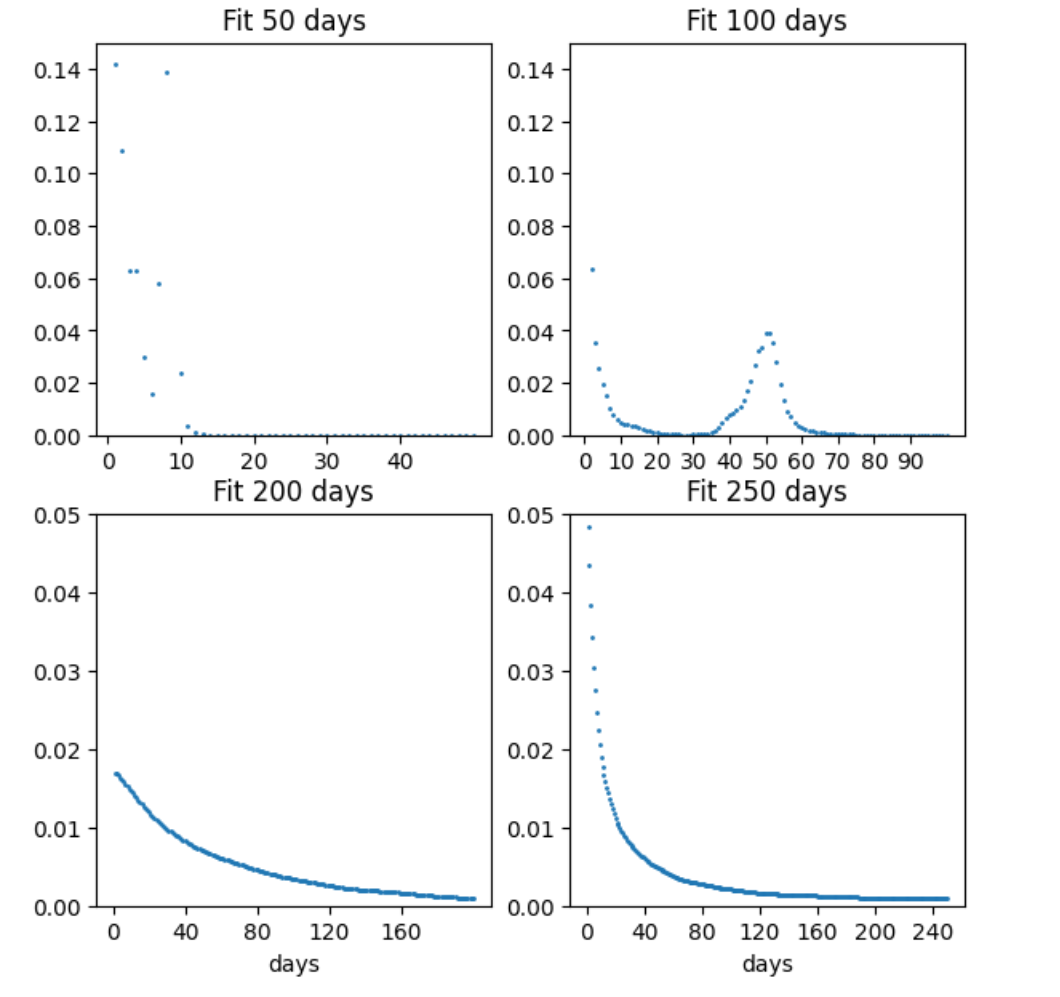}
    \caption{Optimised weights. Own graphics.}
    \label{cloudoptim}
\end{figure}

We optimize weights for four representative periods - 50, 100, 200, and 250 days. The corresponding results are presented in Figure \ref{cloudoptim}. Optimal weights do not have a clear structure for shorter periods, while for longer periods, they are well-formed.

\begin{figure}[!h]
    \includegraphics[scale=0.65]{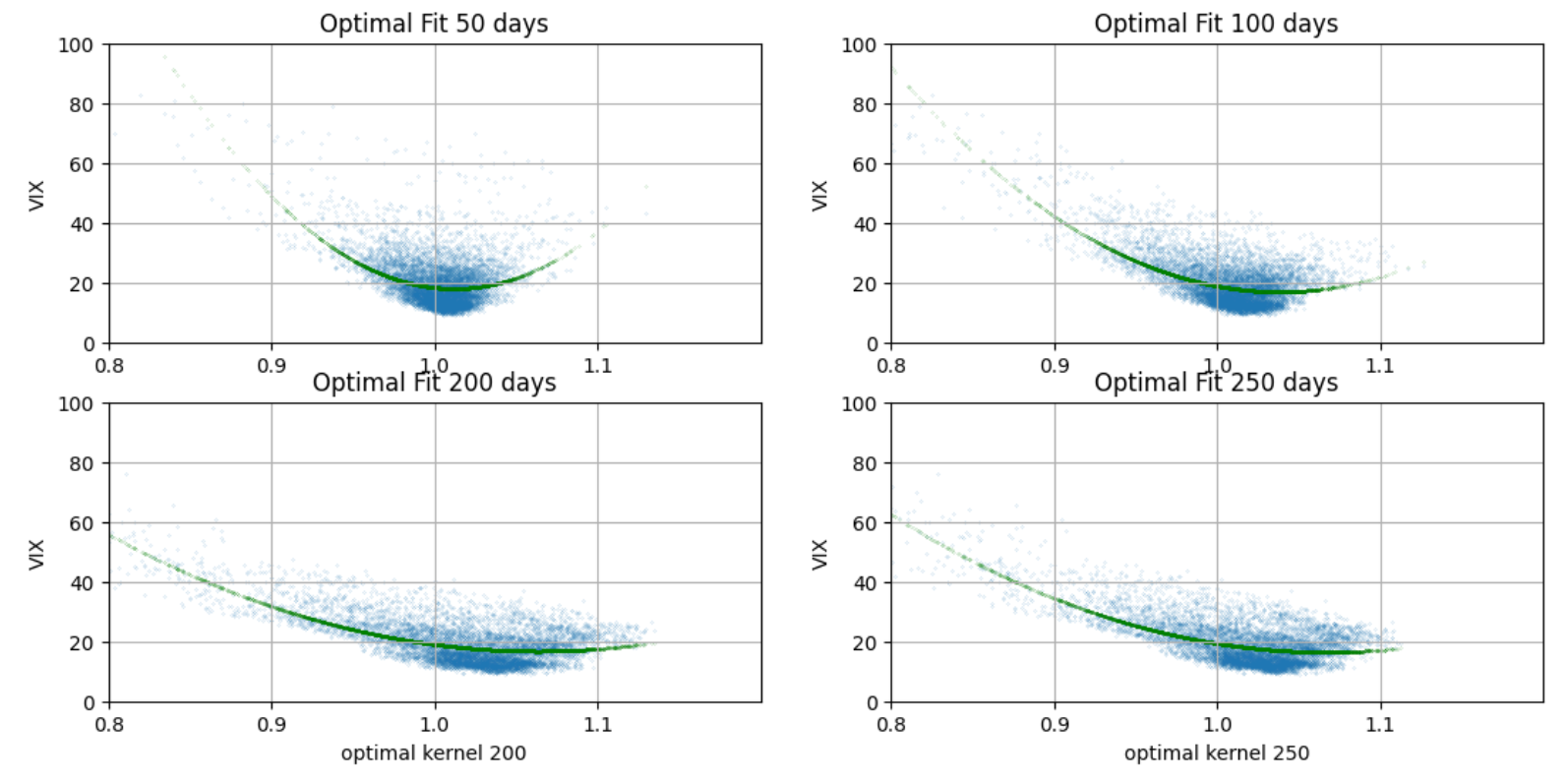}
    \caption{Optimised fit. Own graphics.}
    \label{cloudoptim2}
\end{figure}

\begin{figure}[!h]
    \includegraphics[scale=0.5]{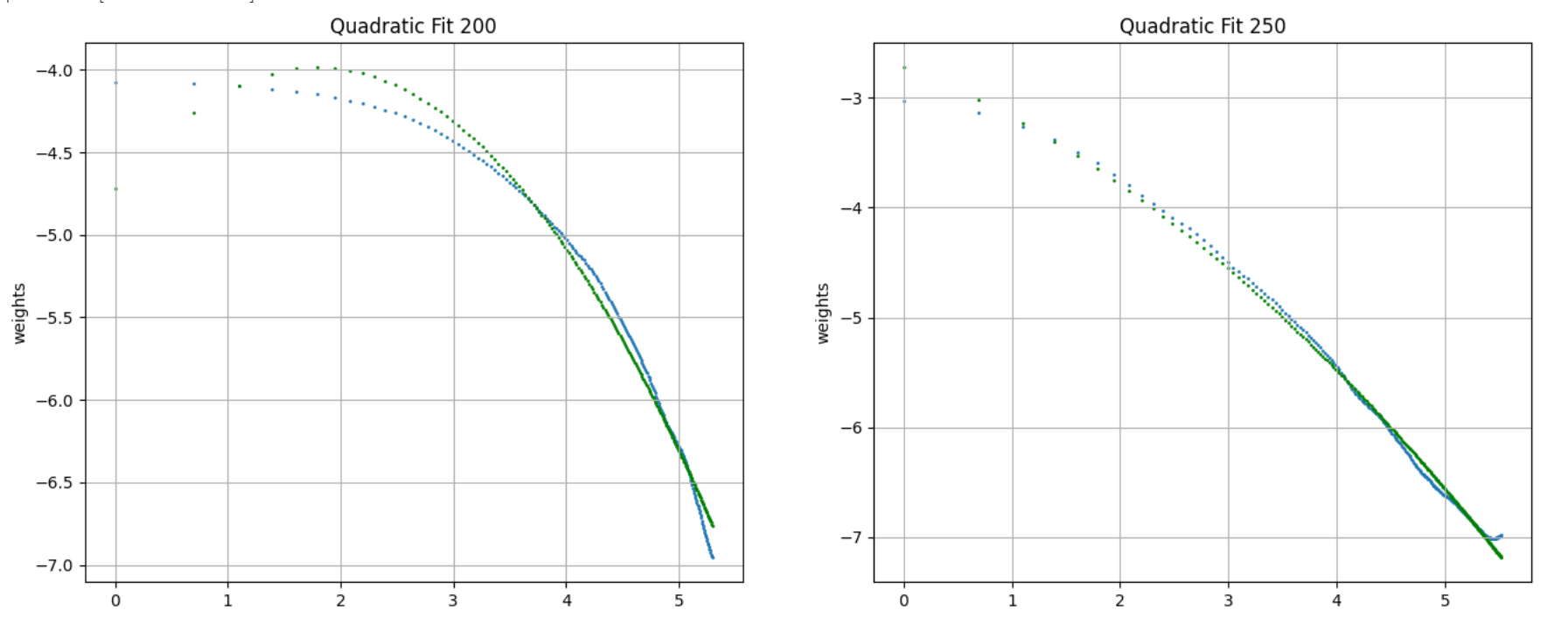}
    \caption{Functional form of the weights. 200 days average: score-weight  0.9948, parameters: [0 -0.38 -0.08]; 250 days average: score-weight  0.9839, parameters: [0 0.82 -0.23]. Own graphics.}
    \label{fig:powerlaw}
\end{figure}

Figure \ref{cloudoptim2} shows that the optimally identified kernel reduces the cloud width compared to the moving average previously calculated. Remarkably, the obtained kernel resembles a power law in the form of a fast-decreasing shape as described \cite{Bergomi16}; see Figure \ref{fig:powerlaw}.

\begin{figure}[!h]
    \centering
    \includegraphics[scale=0.4]{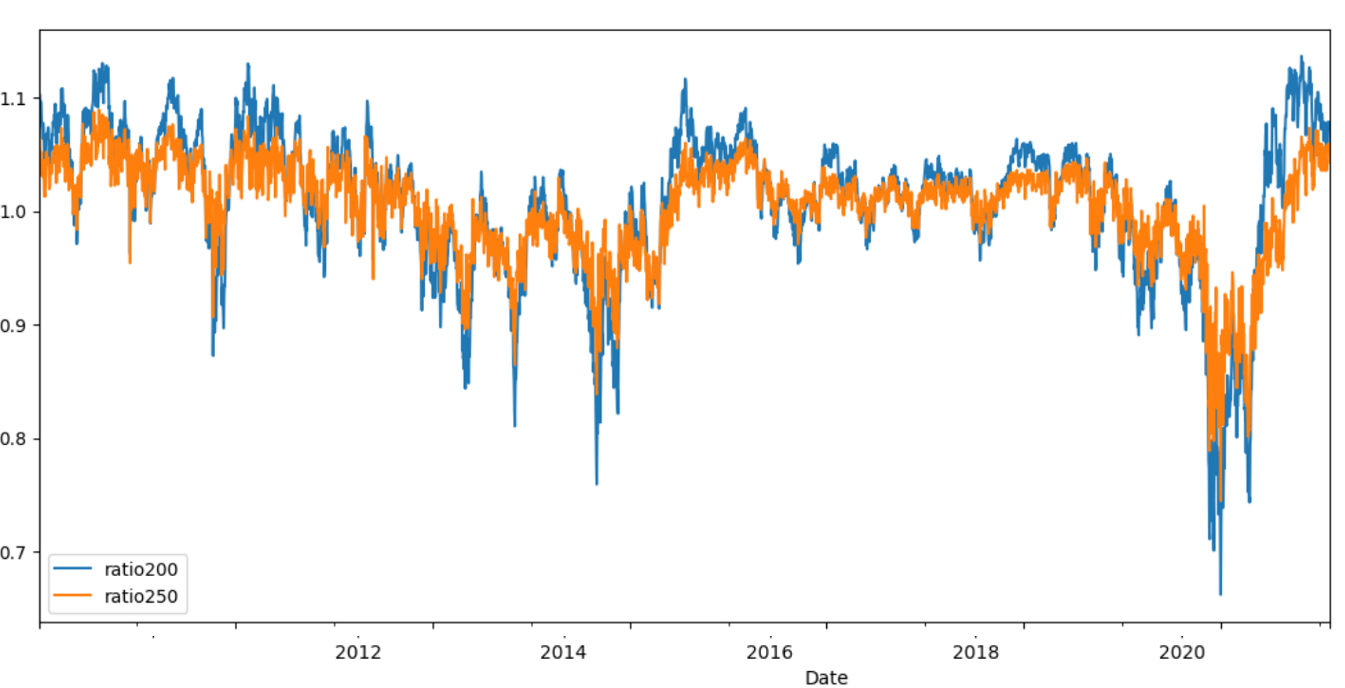}
    \caption{Out Sample : Indices $I_t^{200}$ and $I_t^{250}$. Own graphics.}
    \label{fig:ratio200250}
\end{figure}

The corresponding indices $I_t^{200}$ and $I_t^{250}$ are shown in Figure \ref{fig:ratio200250}.

\subsection{Innovations}

\begin{figure}[!h]
    \centering
    \includegraphics[scale=0.35]{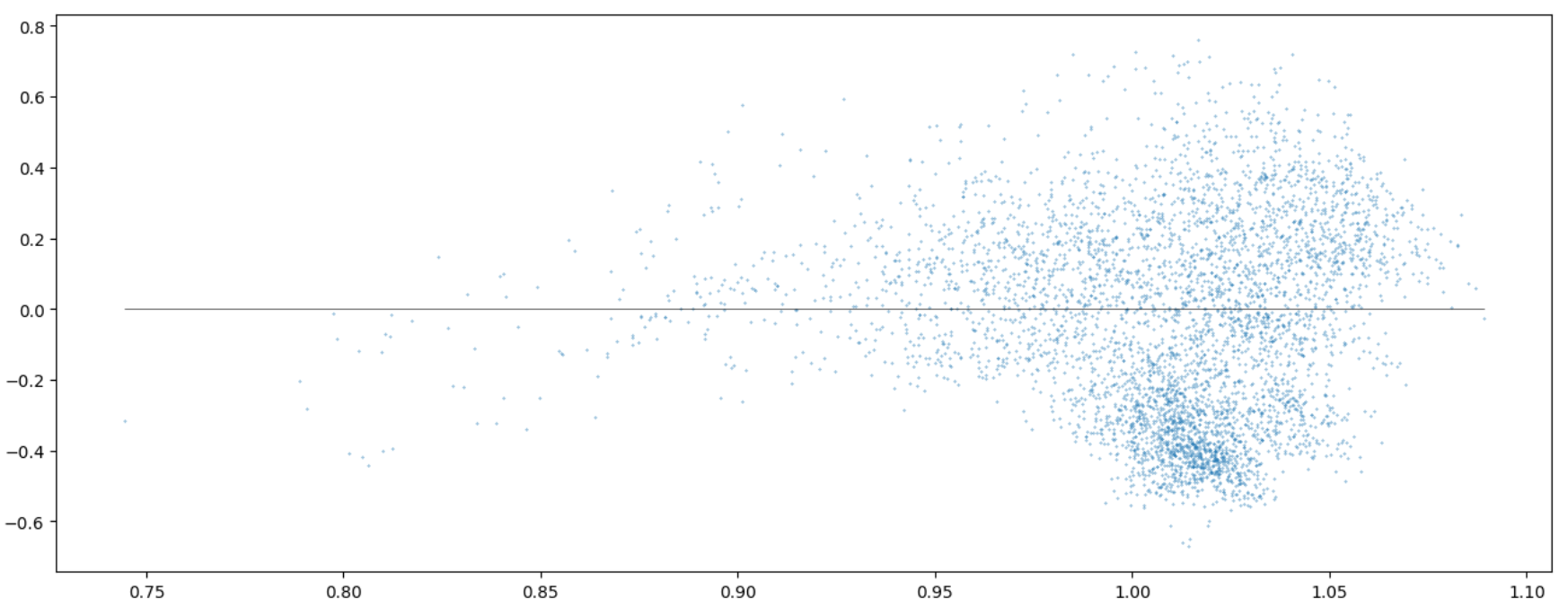}
    \caption{Innovations. Own graphics.}
    \label{fig:innovation}
\end{figure}

The relationship between VIX and SPX is complex and dynamic, and no one-size-fits-all approach works in all market conditions. However, the dimensional approach provides a good proxy for such a relationship. Besides, it gives a causal relationship between levels of VIX and surprises or shocks expressed as the distance between the current SPX spot and its weighted average using the previously identified kernel. In Figure \ref{fig:innovation}, we plot the following ratio:

\begin{equation}
y_t = \ln(\frac{VIX_t}{f(\frac{S_t}{\phi_1 S_{t-n}+...+\phi_nS_{t-1}})}).
\end{equation}
which represents stochastic innovations unexplained by the local volatility contributions.

\begin{figure}[!h]
    \centering
    \includegraphics[scale=0.5]{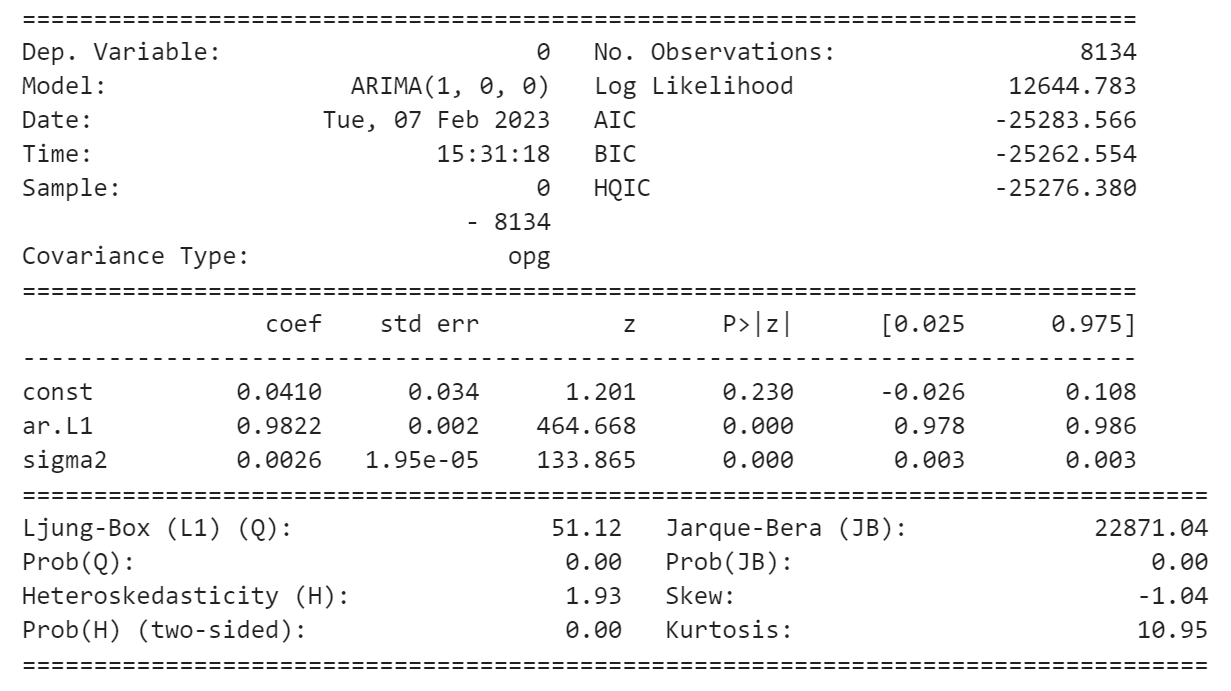}
    \caption{The autoregressive fit of the innovations. Own graphics.}
    \label{fig:my_label}
\end{figure}

We estimate the volatility process using an autoregressive AR(1) the corresponding $y_t$. The fit on $y_t$ is excellent; see Figure \ref{fig:my_label}. It has a $R^2$ score of 96.75\%.

 Thus, the discrete dynamics of the innovations can be written as follows:
\begin{equation}
    y_{t+1} = 0.9822 y_{t} + 0.0026 \epsilon_{t+1},
    \label{eq:orstein}
\end{equation}
where $\epsilon_{t} \sim {\cal{N}}(0,1)$.

\section{Derivatives pricing}

\subsection{Background}

We combine the one-dimensional Monte Carlo simulation and the quantization technique method to design an efficient technique for pricing derivative options on assets in the LSV model. Our approach is similar to the approach proposed in \cite{Lipton22} with few modifications. First, we condition the dynamics on the realization of the $Y$ process. The idea is the same; however, we rely on the functional quantization technique because the conditional price of a derivative is smooth enough with respect to the volatility path so that there is not obligatory to use Monte Carlo on $Y$. In practice, only three quantizers are enough to capture the conditional convexity of the products with respect to this variable. Second, as described in \cite{Lipton22}, we price derivatives conditionally to the $Y$ process. In practice can be done in several ways, depending on the problem's dimensionality. For mono-asset barrier options, one can use PDEs. For multi-dimensional problems, one can use Monte Carlo. It is crucial to note that this pricing methodology is extremely fast and smooth both for the mono and multi assets pricing. Therefore, it is a good candidate for pricing and risk management industrialization.

We start by illustrating the method for vanilla options and consider a call option with maturity $T$ and strike $K$.

We first introduce a functional quantization partition $C_i(y)$ with $Q$ quantizer as described in  \cite{Pages05}. One can note that the Orstein-Uhlenbeck process is treated analytically as an example in \cite{Pages05}.

We define a quantization $\hat{Y}^y$ of $Y$ that can be used in the following quadrature formula. For a given functional $F: L_2[0,T] \mapsto \RR $, and for every $y=(y_1,..., y_Q)$ in $L_2[0,T]^Q$ we get:

\begin{equation}
\EE(F(\hat{Y}^y)) =\sum_{i=1}^{Q} \PP_Y(C_i(y)) F(y).     
\end{equation}

So if one has numerical access to both the Q-quantizer $y$ and its “companion” distribution $\PP_Y(C_i(y))$, the computation is straightforward. 

\begin{eqnarray}
\EE(S_T-K)^+ & = & \EE(\EE((S_T-K)^+|\{Y_t, t\in [0,T]\})\\
     & = & \sum_{i=1}^{Q} \PP_Y(C_i(y)) \EE((S_T-K)^+|y), 
\end{eqnarray}

The process $S$ conditioned on the quantization $y_t$ , that we note $S^y$, is given by:

\begin{eqnarray}
   \frac{d S_t^y}{S_t^y}  & = & \mu^y_t \sigma_{loc} dt + \nu^y_t \sigma_{loc} dB_t, 
   \label{conditional1} \\
    \mu^y_t & = & \frac{\rho}{\nu} (y^{'}_t+\kappa y_t) e^{y_t}, \\ \nu^y_t & = & \sqrt{1-\rho^2}   e^{y_t}.
\end{eqnarray}

Monte Carlo Method can be used to simulate this model. However, the process $S^y$ is no longer a martingale. Instead, it presents a term structure drift and path-dependent volatility.

\subsection{Vanilla option pricing}

In this section, we work on the conditional process (\ref{conditional1}). We can make the pricing using three different approaches:
\begin{itemize}
\item Monte Carlo,
\item Partial differential equation,
\item and Most likely path approach.
\end{itemize}

In the Monte Carlo approach, we discretize the time $t_0=0 < t_1 < ... < t_n=T $ and generate random numbers $\epsilon_i \in {\cal{N}}(0,1)$ i.i.d (independently and identically distributed):

\begin{eqnarray}
   S^y_{0} & = & S_0, \\ 
   S^y_{i+1} & = & (1+(r-q)(t_{i+1}-t_i)) S^y_{i} \\
              & + & S^y_{i} (\mu^y_i \sigma_{loc}(i) (t_{i+1}-t_i) + \nu^y_i \sigma_{loc}(i) \epsilon_i \sqrt{t_{i+1}-t_i} ), \\
   \sigma_{loc}(i) & = & f(\frac{S^y_{i}}{e^{-\kappa t_i} \sum_{j=0}^i e^{\kappa t_j} S^y_{j}}).
\end{eqnarray}
To avoid cumbersome computations, we introduce an additional variable $m$ representing the running mean:
\begin{eqnarray}
   m_i & = &  \sum_{j=0}^i e^{\kappa t_j} S^{y_j}, \\
   m_{0} & = & e^{\kappa t_0} S_0, \\ 
   m_{i+1} & = & m_{i} + e^{\kappa t_{i+1}} S_{i+1}.
\end{eqnarray}

Regarding the partial differential equation (PDE), we introduce a new variable for the weighted average $m_t$:

\begin{equation}
    dm_t = e^{kt}S_t dt,
\end{equation}
The value function $u(t,s,m)$ for the vanilla option satisfies the following PDE:

\begin{equation}
    \partial_t u  + \mu_t^y s \partial_s u + e^{kt} s \partial_m u + 
    \frac{1}{2} (\frac{s}{m})^2 \partial_{ss} u   =  0, 
\end{equation}
\begin{equation}
    u(T,s,m) = (s-K)^+.    
\end{equation}

\begin{figure}[!h]
    \includegraphics[scale=1.1]{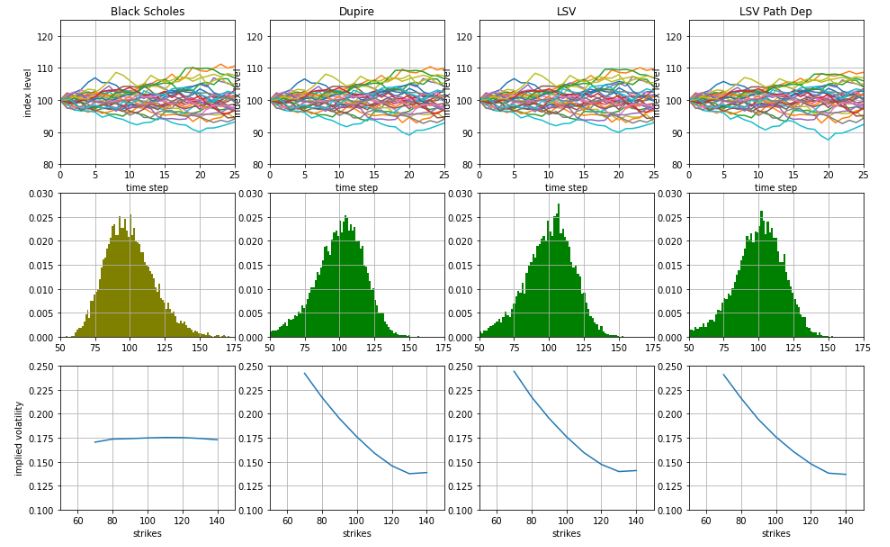}
    \caption{Implied volatilities calculated by several complementary methods. Own graphics.}
    \label{fig:calibration}
\end{figure}

The most likely path, cf \cite{Reghai12}, is also an efficient technique to compute the prices and is well adapted to quantizing the volatility. Figure \ref{fig:calibration} demonstrates the corresponding results 

\subsection{Path-dependent option pricing}

We now have several models calibrated on the same smile, and we shall price different derivative products that depend on the dynamic of the smile; see \cite{Monciaud21} for further details. Indeed, an emblematic product widely used by investors is the up-and-out call. The fact that its gamma exposure changes sign requires the smile to be considered in pricing and hedging \cite{Reghai15}. In addition, the fact that the product can terminate earlier than its maturity if we touch the deactivating barrier means we must unwind our vanilla hedging. The cost of such an operation is also dependent on the smile in the future. Thus the hedging of this product requires taking into account the dynamics of the smile. We compare the different hedging costs using the approach described in \cite{Monciaud21} and the newly calibrated model; see Table \ref{tab:pricing}.

\begin{table}[!h]
    \centering
    \begin{tabular}{|c|c|}
    \hline
       Model & Price \\
       \hline 
       Black Scholes  & $1.93 \pm 2.33 \times 0.04$\\
       \hline
       Local Volatility   &   $2.91   \pm 2.33 \times 0.05$\\
       \hline
       LSV as in \cite{Monciaud21} &  $2.77 \pm 2.33 \times 0.05$\\
       \hline
       LSV PD  & $3.12 \pm 2.33 \times 0.05$\\
       \hline
    \end{tabular}
    \caption{Price Call Up \& Out K=100,B=120.}
    \label{tab:pricing}
\end{table}

Table \ref{tab:pricing2} shows hedging costs for variance and volatility swaps.
\begin{table}[!h]
    \centering
    \begin{tabular}{|c|c|c|}
        \hline
       Model & Variance Swap & Volatility Swap \\
       \hline 
       Black Scholes  & $17.49 \pm 2.33 \times 0.02$ & $17.47 \pm 2.33 \times 0.01$\\
       \hline
       Local Volatility   &   $18.97 \pm 2.33 \times 0.04$ & $18.47 \pm 2.33 \times 0.04$\\
       \hline
       LSV as in \cite{Monciaud21} &  $18.97 \pm 2.33 \times 0.02$ & $18.17 \pm 2.33 \times 0.005$\\
       \hline
       LSV PD  & $18.97 \pm 2.33 \times 0.03$ & $18.39 \pm 2.33 \times 0.05$\\
       \hline
    \end{tabular}
    \caption{Volatility derivatives prices.}
    \label{tab:pricing2}
\end{table}

\subsection{Final set of equations }

Using the parametrization in equation (\ref{eq:orstein}) above, as well as a zero correlation between the spot and the volatility, we get the following set of equations:

\bea
    \frac{d S_{t}}{S_{t}} &=& (r-q)dt + \sigma_{loc}  e^{Y_t} dB_t, \\
    dY_t & = & -{\kappa} Y_t dt + \nu dW_t, \\
     \frac{dI_t}{I_t} &=& ((r-q)-\phi(t,t)I_t) dt + \sigma_{loc}  e^{Y_t} dB_t.
\eea

\bea
\sigma_{loc} = f\left(\frac{S_t}{ \int_{-\infty}^t \phi(t,u) S_udu}\right).
\eea
where $r,q$ are the risk neutral rate and dividend yield and $\phi(t,u)$ is the power law weight function given in Figure \ref{fig:powerlaw}.

\section{Scenario generation}

Generating consistent trajectories for VIX and SPX is essential for both the buy and sell sides. For the buy side, it is crucial to generate realistic paths to test strategies. On the sell side, it is the question of consistency and having a model matching both SPX and Vix smile.

\section{Conclusion}

In conclusion, dimensionless data modeling has proven valuable in understanding the complex relationship between VIX and SPX. By removing the units in the SPX and putting some knowledge about how markets function, we could identify the proper kernel for modeling the link between SPX and VIX.

After studying the residuals, we could identify that the log ratio is well represented with a fast mean reverting process. 

Combining these two techniques, we end up with a parsimonious generating model that captures with fidelity some of the main features in financial markets.

Overall, the use of dimensionless data modeling in the analysis of VIX and SPX highlights the importance of data normalization and the benefits of using this technique to uncover patterns and relationships in complex data sets. 


\end{document}